\documentclass[12pt]{article}
 
\usepackage{graphicx}
\usepackage{amsfonts}
\usepackage{amsthm}
\usepackage[top=2.75cm, bottom=2.75cm, left=2.75cm, right=2.75cm]{geometry} 

\newtheorem{proposition}{Proposition}

\newtheorem{definition}[proposition]{Definition}

\newtheorem*{theorem}{Theorem}

\newcommand{\coker}{\mathrm{coker}}
 
\newcommand{\beq}{\begin{equation}} 
\newcommand{\eeq}[1]{\label{#1}\end{equation}}
\newcommand{\ber}{\begin{eqnarray}} 
\newcommand{\eer}[1]{\label{#1}\end{eqnarray}}

\begin{document}
\renewcommand{\theequation}{\thesection.\arabic{equation}}  
\setcounter{page}{0}

\thispagestyle{empty}

\begin{flushright} \small
UUITP-04/07 \\ HIP-2007-12/TH \\  YITP-SB-07-09 \\ 
\end{flushright}
\smallskip
\begin{center} \LARGE
{\bf A potential for Generalized K\"ahler Geometry}
 \\[12mm] \normalsize
{\bf Ulf~Lindstr\"om$^{a,b}$, Martin Ro\v cek$^{c}$, 
Rikard von Unge$^{d}$, and Maxim Zabzine$^{a}$} \\[8mm]
{\small\it
$^a$Department of Theoretical Physics 
Uppsala University, \\ Box 803, SE-751 08 Uppsala, Sweden \\
~\\
$^b$HIP-Helsinki Institute of Physics, University of Helsinki,\\
P.O. Box 64 FIN-00014  Suomi-Finland\\
~\\
$^c$C.N.Yang Institute for Theoretical Physics, Stony Brook University, \\
Stony Brook, NY 11794-3840,USA\\
~\\
$^{d}$Institute for Theoretical Physics, Masaryk University, \\ 
61137 Brno, Czech Republic \\~\\}
\end{center}
\vspace{10mm}
\centerline{\bfseries Abstract} 
\bigskip
\noindent  
We show that, locally, all geometric objects of Generalized K\"ahler Geometry 
can be derived from a function $K$, the ``generalized K\"ahler potential''.
The metric $g$ and two-form $B$ are determined as nonlinear functions
of second derivatives of $K$. These nonlinearities are shown to arise via a 
quotient construction from an auxiliary local product (ALP) space.
\eject
\normalsize

%\addtocontents{toc}
%\tableofcontents

%\end{titlepage}

\eject

\section{Introduction}
Generalized K\"ahler Geometry  \cite{gualtieri} is a special and
particularly interesting example of Generalized Geometry 
\cite{hitchinCY,gualtieri}. It succinctly encodes the bihermitean
geometry of Gates, Hull and Ro\v cek \cite{Gates:1984nk} in
terms of a pair of commuting generalized complex structures
$({\cal{J}}_1,{\cal{J}}_2)$. Here we use the equivalent 
bihermitean data to give a complete (local)
description of Generalized K\"ahler Geometry. We show that
there exists a function $K$ with the interpretation of a
generating function for symplectomorphisms between two sets
of coordinates in terms of which all geometric quantities
may be described. In particular, the combination
$E={\textstyle{1\over 2}} (g+B)$ of the metric and antisymmetric $B$-field
is given as a nonlinear expression in terms of second
derivatives of $K$. When further analyzed, this expression
has the structure of a quotient from some higher dimensional
space. We find this auxiliary space, which we call an ALP
space, and display the quotient structure. Finding a
Generalized K\"ahler Geometry is thus reduced to a linear
construction in terms of the corresponding $K$ on the ALP.

The bihermitean geometry was first derived as the target space
geometry of ${\cal{N}}=(2,2)$ supersymmetric nonlinear sigma
models \cite{Gates:1984nk}. Correspondingly, 
all our constructions have a sigma
model realization, and many of the proofs were derived in
that setting. In particular, the 
``generalized K\"ahler potential'' $K$ has an interpretation as 
the superspace sigma model Lagrangian\footnote{A similar
situation arises in the projective superspace description of
hyperk\"ahler sigma models \cite{LR}.}.
Detailed descriptions of this approach may be found
in the two articles on which the presentation is based
\cite{Lindstrom:2005zr,Lindstrom:2007qf}.
 
\section{Generalized Complex Geometry (GCG)}
\label{OFF}

This section contains a brief recapitulation of the
salient features of GCG. The references for the mathematical
aspects are chiefly \cite{hitchinCY,gualtieri}, while
a presentation more accessible to physicists (containing,
{\it e.g.,} coordinate and matrix descriptions) may be found in
\cite{Lindstrom:2004iw,Zabzine:2006uz}.

A  generalized almost complex structure is an algebraic 
structure on the sum of the tangent and cotangent 
bundles of a manifold ${\cal{M}}$ defined via
\beq
{\cal{J}}\in {\rm End}(T\oplus T^*) 
\eeq{7}
such that 
\beq
{\cal{J}}^2=-1
\eeq{8}
and
\beq
{\cal{J}}^t{\cal{I}}{\cal{J}}={\cal{I}}
\eeq{9}
where ${\cal{I}}$ is the metric corresponding to the 
natural pairing of elements $X+\eta,Y+\lambda \in T\oplus T^*$:
\beq
<X+\eta,Y+\lambda>={\textstyle{1\over 2}}  (\imath_X\lambda+\imath_Y\eta)~.
\eeq{90}
The generalized almost complex structure is a (twisted) 
generalized complex structure {\it iff} it is integrable 
with respect to the (twisted) Courant bracket
\beq
[X+\eta,Y+\lambda]_C:=[X,Y]_L+{\cal{L}}_X 
\lambda-{\cal{L}}_Y \eta-{\textstyle{1\over 2}}  d( \imath_X
\lambda-\imath_Y\eta) +\imath_X\imath_Y H
\eeq{10}
where $[X,Y]_L$ is the Lie-bracket and 
``twisted'' refers to inclusion of the last term 
involving the closed three-form $H$.
Integrability is defined by the requirement 
that the distributions defined by the projection operators
\beq
{\textstyle{1\over 2}} \left(1\pm i{\cal{J}}\right)
\eeq{110}
are involutive with respect to the Courant-bracket.
Of great importance to physical applications is the 
fact that the automorphisms of the Courant-bracket, 
in addition to diffeomorphisms, also contain the 
$b$-transform:
\beq
e^b(X+\eta)=X+\eta+\imath_X b
\eeq{11}
where $b$ is a closed two-form: $db=0$. 
When acting on a generalized complex structure ${\cal{J}}$
it produces an equivalent generalized complex 
structure ${\cal{J}}_b$. In a matrix representation this reads;
\beq
{\cal{J}}_b=\left(\begin{array}{cc}
1&0\cr
b&1\end{array}\right){\cal{J}}\left(\begin{array}{cc}
1&0\cr
-b&1\end{array}\right)
\eeq{12}
We now restrict our attention to a subset of  
generalized complex geometries: generalized K\"ahler geometries (GKG).
These are GCG's with two commuting generalized 
complex structures ${\cal{J}}_{1,2}$
whose product defines a positive definite metric ${\cal{G}}$  
on $T\oplus T^*$ that squares to the identity 
\cite{gualtieri};
\ber
&&[{\cal{J}}_{1},{\cal{J}}_{2}]=0\cr
&&\cr
&& {\cal{G}} =-{\cal{J}}_{1}{\cal{J}}_{2}\cr
&&\cr
&&{\cal{G}}^2=1
\eer{13}
This is the proper setting for the bihermitean geometry of 
\cite{Gates:1984nk}, which we now describe.

\section{Bihermitean Geometry}

The data found in \cite{Gates:1984nk} to be necessary and
sufficient to describe the target space geometry of a
supersymmetric nonlinear sigma model with $(2,2)$
supersymmetry consists of two (ordinary) complex structures
$J_+$ and $J_-$, a metric $g$, hermitean with respect to
both complex structures, and a closed three-form $H$ with a
(local) two-form potential $B$.

The structures $(J_\pm,g,B)$ satisfy
\ber
&& J^2_{\pm}=-1\cr
&&\cr
&&J_{\pm}^t g  J_{\pm}=g\cr
&& \cr
&&\nabla^{(\pm)} J_{\pm}=0\cr
&&\cr
&&H = d_+^c \omega_+ = - d_-^c \omega_-
\eer{bla}
where $\omega_\pm$ are the two-forms $gJ_{\pm}$,
$d^c_\pm$ is $d^c$ with respect to $J_\pm$ respectively, 
and the covariant derivatives have (Bismut) torsion \cite{Yano}
\beq
\nabla^{(\pm)}:=\nabla\pm g^{-1}H, \quad dH=0~,
\eeq{3}
where $\nabla$ is the Levi-Civita connection.

When $J_\pm$ commute, it was found in \cite{Gates:1984nk} that 
there exist coordinates $z,\bar z$ and $z', \bar z'$ that coordinatize 
$\ker(J_+\mp J_-)$, respectively, and for which the metric and 
three-form are given as derivatives of a function
$K(z,\bar z, z, \bar z')$. Explicitly:
\ber
\omega_\pm=(d^c_\mp d+d^c_\Pi d^c_\pm) K~& \Leftrightarrow&~
g_{z\bar z}=\partial_z\partial_{\bar z}K~,~~
g_{z'\bar z'}=-\partial_{z'}\partial_{\bar z'}K~,\cr\cr
H=dd^c_+d^c_-K~&\Leftrightarrow&~
H_{z,\bar z,z'}=\partial_z\partial_{\bar z}\partial_{z'} 
K~,~~ H_{z,\bar z,\bar z'}=-\partial_z\partial_{\bar z}\partial_{\bar z'}K~,\cr\cr
&&H_{z',\bar z',z}=-\partial_z'\partial_{\bar z'}\partial_z 
K~,~~ H_{z',\bar z',\bar z}=\partial_{z'}\partial_{\bar z'}\partial_{\bar z}K~
\eer{6}
where $d^c_\Pi$ is the $d^c$ operator defined with respect 
to the product $\Pi:=J_+J_-$ and the differentials 
$d^c_+,d^c_-,d^c_\Pi$ anticommute.

A complete description for geometries with a nonvanishing $\coker [J_+, J_-]$ 
was an open problem until our paper
\cite{Lindstrom:2005zr}. Before describing the solution 
to the problem in sections \ref{COKER} and \ref{gencase}, 
we recall Gualtieri's map of the bihermitean data $(J_\pm,g,B)$ into 
Generalized K\"ahler Geometry \cite{gualtieri}:
\beq
{\cal{J}}_{1,2}=\left(\begin{array}{cc}
1&0\cr
B&1\end{array}\right)\left(\begin{array}{cc}
J_+\pm J_-&-(\omega_+^{-1}\mp\omega_-^{-1})\cr
\omega_+\mp\omega_-&-(J_+^t\pm J_-^t)\end{array}\right)
\left(\begin{array}{cc}
1&0\cr
-B&1\end{array}\right)~.
\eeq{14}
\beq
 {\cal{G}}=\left(\begin{array}{cc}
1&0\cr
B&1\end{array}\right)
\left(\begin{array}{cc}
0&g^{-1}\cr
g&0\end{array}\right)
\left(\begin{array}{cc}
1&0\cr
-B&1\end{array}\right)
\eeq{15}
Note that what looks like a $b$-transform is not, since $dB=H\ne0$.

It is an interesting fact that if we introduce the 
usual GCS's corresponding to $J_\pm$;
\beq
{\cal{J}}_\pm=\left(\begin{array}{cc}
J_\pm&0\cr
0&-J_\pm^t\end{array}\right)
\eeq{140}
the map (\ref{14}) may be summarized as\footnote{This 
structure is related to how the map is derived in chapter 6 of \cite{gualtieri}}
\beq
{\cal{J}}_{1,2}=\Pi_+{\cal{J}}_{+B}\pm\Pi_-{\cal{J}}_{-B}~,
\eeq{150}
with
\beq
\Pi_\pm:={\textstyle{1\over 2}} \left(1\pm{\cal{G}}\right) ~.
\eeq{160}

\section{Poisson-structures}
\label{Poise}

There are three Poisson-structures relevant to our discussion. 
We first consider two real ones:
\beq
\pi_\pm :=(J_+\pm J_-)g^{-1}=-g^{-1}(J_+\pm J_-)^t~.
\eeq{19}
These were introduced in \cite{Lyakhovich:2002kc}, 
where they were used essentially as described below, 
simplifying an earlier derivation in \cite{Ivanov:1994ec}. 
They ensure the existence of coordinates adapted to 
$\ker (J_+ - J_-) \oplus \ker (J_+ + J_-)$.

In a neighborhood of a regular point $x_0$ of  $\pi_-$ 
we may choose coordinates $x^A$ whose tangents lie in the kernel of $\pi_-$:
\beq
\pi_-^{A\mu}=0, \Rightarrow J^A_{+\mu}=J^A_{-\mu}~.
\eeq{190}
Similarily, in a neighborhood of a regular point of  $\pi_+$ 
we have coordinates $x^{A'}$ such that
\beq
\pi_+^{A'\mu}=0, \Rightarrow J^{A'}_{+\mu}=-J^{A'}_{-\mu}
\eeq{20}
It then follows from the nondegeneracy of $\pi_+\pm\pi_-$ 
that the Poisson brackets defined by
$\pi_+$ and by $\pi_-$ cannot have common Casimir 
functions. In other words, the directions 
$A$ and $A'$ cannot coincide. The result is that we may write
\beq
J_\pm = \left( \begin{array}{cccc}
* & * &  * &  *\\
**&*& *& *\\
0 & 0 & I_c & 0\\
0 & 0 & 0& \pm I_t
\end{array} \right)~,
\eeq{speccorosl278}
in coordinates adapted to
\beq
\ker (J_+ - J_-) \oplus \ker (J_+ + J_-) \oplus  \coker [J_+, J_-]~,
\eeq{decompjJJJ}
where we write a canonical complex structure as
\beq
I=\left(\begin{array}{cc}
i&0\cr
0&-i\end{array}\right)
\eeq{111}
and $c$ labels the $A$ and $t$ labels the  $A'$ directions.

A third Poisson structure $\sigma$, related to the 
real Poisson structures, was introduced in \cite{hitchinP}.
\beq
\sigma :=  [J_+, J_-] g^{-1}= \pm (J_+ \mp J_-) \pi_\pm=\mp(J_+\pm J_-)\pi_\mp~.
\eeq{definHitchpoisos}
The relation to $\pi_\pm$ implies that
\beq
\ker \sigma = \ker \pi_+ \oplus \ker \pi_-~.
\eeq{kernelsporisoap}
Tangents to the symplectic leaf for $\sigma$ 
lie in $\coker [J_+, J_-]$. Using (\ref{kernelsporisoap}), 
we can use $\sigma$ to investigate the remaining 
directions in (\ref{speccorosl278}). We need that
\beq
J_\pm \sigma J_\pm^t=-\sigma\ 
\Rightarrow\sigma = \sigma^{(2,0)} + \bar{\sigma}^{(0,2)}~,
\eeq{sigamdecomshdol}
with the decomposition with respect to 
{\em both} $J_+$ and $J_-$. Further $\sigma^{(2,0)}$ is holomorphic
\cite{hitchinP}:
\beq
\bar\partial \sigma^{(2,0)}=0~,
\eeq{defindholomr}
We now investigate the structure of the cokernel.

\section{The structure of $\coker [J_+, J_-]$}
\label{COKER}

It is convenient to first treat the case $\ker [J_+, J_-] =\emptyset$. 
Then we have at our disposal a symplectic form
$\Omega$:
\beq
\Omega := \sigma ^{-1}~,\quad d\Omega=0, \quad  
J^t_\pm\Omega J_\pm=-\Omega~.
\eeq{deftwofrom}
Choosing complex coordinates with respect to $J_+$,
\beq
J_+ = \left( \begin{array}{cc}
I_s & 0\\
0 & I_s
\end{array} \right),
\eeq{form23-a}
with $I$ as in (\ref{111}), we have the decomposition
\beq
\Omega = \Omega_+^{(2,0)} + \bar{\Omega}_+^{(0,2)}~.
\eeq{decompomega}
and
\beq
\partial \Omega_+^{(2,0)}=0~,\qquad\bar{\partial} \Omega_+^{(2,0)}=0~,
\eeq{cloholomomega}
which means in particular that $\Omega_+^{(2,0)}$ is holomorphic. 
We may then choose Darboux coordinates such that
\beq
\Omega_+^{(2,0)} = d q^a \wedge d p^{a}~,\qquad
\bar{\Omega}_+^{(0,2)} = d\bar q^{\bar{a}} \wedge d \bar p^{\bar a }~ .
\eeq{drholomcoord}

Alternatively, we can choose complex coordinates with respect  
to $J_-$ and similarily derive:
\beq
\Omega_-^{(2,0)} = d Q^{a'} \wedge d P^{a'}~,\qquad
\bar{\Omega}_-^{(0,2)} = d\bar Q^{\bar a'} \wedge d\bar P^{\bar a'}~.
\eeq{complckdoe22}

The transformation $\{q,p\}\to\{Q,P\}$  is a canonical transformations
(symplectomorphism) and can thus be specified by a generating 
function\footnote{There always exists at least one
polarization such that any symplectomorphism,
can be written in terms of such a generating function \cite{arnold}. } 
$K(q,P)$. Thus in a neighborhood, the canonical transformation is
given by the generating function $K(q,P)$
\beq
p = \frac{\partial K}{\partial q}~,\qquad Q = \frac{\partial K}{\partial P}~.
\eeq{denegaiwp298}
We calculate all our geometic structures $J_+$, $J_-$, 
$\Omega$ and $g$ in the ``mixed'' coordinates $\{q, P\}$, 
using the transformation matrices
\beq
\frac {\partial (q, p)}{\partial (q, P)}  = \left( \begin{array}{cc}
1 &  0 \\
\frac{\partial p}{\partial q} & \frac{\partial p} {\partial P}
\end{array}
\right)  = \left( \begin{array}{cc}
1 &  0 \\
\frac{\partial^2 K}{\partial q \partial q} & \frac{\partial^2 K} {\partial P\partial q}
\end{array}
\right) := \left( \begin{array}{cc}
1 & 0 \\
K_{LL} & K_{LR}
\end{array}
\right)
\eeq{tarsnforamslwo}
and
\beq
\frac {\partial (Q, P)}{\partial (q, P)}  = \left( \begin{array}{cc}
\frac{\partial Q}{\partial q} & \frac{\partial Q} {\partial P} \\
0 & 1
\end{array}
\right)  = \left( \begin{array}{cc}
\frac{\partial^2 K}{\partial q \partial P} & \frac{\partial^2 K} {\partial P\partial P}\\
0 &  1
\end{array}
\right) := \left( \begin{array}{cc}
K_{RL}& K_{RR}\\
\,\,0 & \,\,1
\end{array}
\right) ,
\eeq{tarsnforamslwonew}
where the labels $L$ and $R$ are shorthand for the
coordinates $\{q^a,\bar q^{\bar a}\}$ and $\{P^{a'},\bar
P^{\bar a'}\}$ in (\ref{drholomcoord}) and
(\ref{complckdoe22}) above. The expressions for $J_\pm$ are
nonlinear and can be read off from the formulae given below 
for the general case with $\ker [J_+, J_-] \ne\emptyset$.
The symplectic structure $\Omega$ is the only linear
function of $K$
\beq
\Omega = K_{AA'}d q^{A} \wedge d P^{A'}~,
\eeq{omegachg}
where we use the collective notation $A=\{a,\bar a\}$, 
$A'=\{a',\bar a'\}$. We write  $\Omega$ as a matrix
\beq
\Omega = \left(\begin{array}{cc}
0 & K_{LR} \\
- K_{RL} & 0
\end{array} \right).
\eeq{defkspoap}
and  find the metric and $B$-field using this 
matrix as follows \cite{Bogaerts:1999jc} 
({\it cf.} \ref{definHitchpoisos})
\beq
g = \Omega [J_+, J_-]~, \quad B=\Omega \{J_+, J_-\}~.
\eeq{definsmetric}

\section{The general case}
\label{gencase}

In the general case when both $\ker [J_+, J_-]$ and 
$\coker [J_+, J_-]$ are nonempty, 
we combine the discussion leading to (\ref{speccorosl278}) 
in section \ref{Poise} 
with that of the previous section. 
The formulas that we compute are rather complicated, and we
have not found a suitable coordinate free way to express them.

We assume that in a neighborhood of $x_0$, the ranks of $\pi_\pm$ are  
constant, and as result, the rank of $\sigma$ is constant.  We   
work in coordinates $\{q,p,z,z'\}$ adapted to the symplectic 
foliation of $\sigma$ as well as to the description of 
$\ker [J_+, J_-]$ given in section \ref{Poise}.
In  such coordinates
$J_+$ may be taken to have the canonical 
form\footnote{For historical reasons related to the 
origin in sigma models involving chiral and twisted 
chiral superfields, we again use the labels 
$c$ and $t$ for the $z$ and $z'$ directions.}
\beq
J_+ = \left(\begin{array}{cccc}
I_s & 0 & 0& 0 \\
0 & I_s  & 0 & 0\\
0 & 0 & I_c & 0\\
0 & 0 & 0 &  I_t
\end{array} \right),
\eeq{canoniicalfromforJ}
where $\{q,p\}$ are Darboux coordinates for a symplectic
leaf of $\sigma$ and  $\{z,z'\}$ parametrize the kernels of
$\pi_\mp$. Similarily, there are coordinates $\{Q,P,z,z'\}$
where $J_-$ takes a diagonal form (identical to that of
$J_+$ in (\ref{canoniicalfromforJ}) except for a change of
sign in the last entry; $I_t\to -I_t$). For every leaf
separately we may now apply the arguments of section
\ref{COKER}. There thus exists a generating function
$K(q,P,z,z')$ for the symplectomorphisms between the two
sets of coordinates. The transformation matrices to the
coordinates $\{q,P,z,z'\}$ are given by the obvious
extensions of (\ref {tarsnforamslwo}) and (\ref
{tarsnforamslwonew}) to include $z,z'$. The expression for
$J_\pm$ in the ``mixed'' coordinates are then\footnote{The rows and columns of this
matrix correspond to the directions along  $\{q,P,z,z'\}$.}:
\beq
J_+ = \left( \begin{array}{cccc}
I_s & 0 & 0 &  0\\
K^{-1}_{RL}C_{LL} & K_{RL}^{-1}I_s K_{LR} &  K_{RL}^{-1}C_{Lc} &
K_{RL}^{-1}C_{Lt} \\
0 & 0 & I_c & 0 \\
0 &0 & 0 & I_t
\end{array} \right),
\eeq{fullJallcasesl}
and
\beq
J_-= \left( \begin{array}{cccc}
K_{LR}^{-1}I_s K_{RL}  & K_{LR}^{-1}C_{RR} & - K_{LR}^{-1}C_{Rc} &
K_{LR}^{-1}A_{Rt} \\
0 &  -I_s & 0 & 0\\
0 & 0 & I_c & 0 \\
0 &0 & 0 & -I_t
\end{array} \right),
\eeq{fullJallcaseslextra}
where
\ber
K_{LR}^{-1} &=& (K_{RL})^{-1}~,\cr
C &=& IK-KI
=\left(\begin{array}{cc}
0 & 2i K\cr
-2i K & 0
\end{array}\right),\cr\cr\cr
A &=& IK+KI
=\left(\begin{array}{cc}
2i K & 0\cr
0 & -2i K
\end{array}\right),
\eer{commnotation}
where we suppress the indices in the last two 
entries\footnote{In (\ref{commnotation}),
the rows and columns correspond to $\{A,\bar A\}$, 
where $\{A\}$ are any one of the $\{q,P,z,z'\}$ 
directions in (\ref{fullJallcasesl}),(\ref{fullJallcaseslextra}),(\ref{E}).}.
The relations (\ref{definsmetric}) no-longer hold when
$\ker[J_+, J_-]\ne \emptyset$. The definition (\ref{definHitchpoisos}) 
of the Poisson structure $\sigma$ still determines the metric, 
except along the kernel. However, the 
additional relation $H = \pm d^c_{\pm} \omega_\pm$ from (\ref{bla}) 
provides us with an equation for the remaining components of $g$ 
and allows us to find the $B$-field. From the sigma model 
we already know the solution; the 
sum $E={\textstyle{1\over 2}} (g+B)$ of the metric $g$ and 
$B$-field takes on the explicit form:
\ber
E_{LL} &=& C_{LL}K_{LR}^{-1}I_sK_{RL} \cr
E_{LR} &=& I_sK_{LR}I_s + C_{LL}K_{LR}^{-1}C_{RR} \cr
E_{Lc} &=& K_{Lc} + I_s K_{Lc} I_c + C_{LL}K_{LR}^{-1}C_{Rc}\cr
E_{Lt} &=& -K_{Lt} - I_s K_{Lt} I_t + C_{LL}K_{LR}^{-1}A_{Rt}\cr
E_{RL} &=& -K_{RL}I_s K_{LR}^{-1} I_s K_{RL}\cr
E_{RR} &=& -K_{RL}I_s K_{LR}^{-1} C_{RR}\cr
E_{Rc} &=& K_{Rc} - K_{RL}I_s K_{LR}^{-1} C_{Rc}\cr
E_{Rt} &=& -K_{Rt} - K_{RL}I_s K_{LR}^{-1} A_{Rt}\cr
E_{cL} &=& C_{cL}K_{LR}^{-1}I_s K_{RL}\cr
E_{cR} &=& I_c K_{cR} I_s + C_{cL}K_{LR}^{-1}C_{RR}\cr
E_{cc} &=& K_{cc}+I_c K_{cc} I_c + C_{cL}K_{LR}^{-1}C_{Rc}\cr
E_{ct} &=& -K_{ct}-I_c K_{ct}I_t + C_{cL}K_{LR}^{-1}A_{Rt}\cr
E_{tL} &=& C_{tL}K_{LR}^{-1}I_s K_{RL}\cr
E_{tR} &=& I_t K_{tR} I_s + C_{tL}K_{LR}^{-1}C_{RR}\cr
E_{tc} &=& K_{tc} + I_t K_{tc} I_c + C_{tL}K_{LR}^{-1}C_{Rc}\cr
E_{tt} &=& -K_{tt} - I_t K_{tt} I_t + C_{tL} K_{LR}^{-1} A_{Rt}
\eer{E}

In view of this last relation (\ref{E}) as well as (\ref{fullJallcasesl}), 
(\ref{fullJallcaseslextra}) and 
(\ref {defkspoap}) it seems appropriate to 
call $K$ the generalized K\"ahler potential.

\section{Linearization of Generalized K\"ahler geometry}
As can be seen in (\ref{E}), the expressions for the metric
and the $B$-field in terms of the generalized K\"ahler
potential are highly nonlinear in contrast to what is the
case in ordinary K\"ahler geometry. The nonlinearities 
all stem from the cokernel
of $[J_+,J_-]$: when $[J_+,J_-]=0$,
the metric and B-field can be expressed as linear functions
of the Hessian of $K$. We note that the structure of the
nonlinearities are such as one might expect from a quotient
construction. One is therefore faced with a natural
question: Is there a space in which all the geometric data
is encoded linearly with respect to the generalized K\"ahler
potential and which gives the nonlinear Generalized K\"ahler
geometry through a quotient construction?

Inspired by superspace sigma-models, we have found the
following local prescription for generating such a space:
In local coordinates, we simply make the substitution
\beq
q\to z_L+z'_L~,~~P\to z_R+\bar z'_R~
\eeq{subsALP}
in the Generalized K\"ahler potential $K$ and use 
the naive commuting complex structures
$J_\pm$ in which the $z$-coordinates are biholomorphic and the 
$z'$-coordinates are $J_+$ holomorphic and $J_-$ antiholomorphic.
This introduces isometries of the metric (\ref{6}) corresponding
to shifts of $z_L,z'_L$ that preserve the sum, and similarly for
$z_R,\bar z_R$; taking a quotient with respect to these
isometries leads to the original nonlinear model.
We now attempt to make mathematical sense of this prescription.
We begin with some definitions:

\begin{definition}[BiLP]
A Generalized K\"ahler Geometry with commuting complex 
structures will be called a Bihermitean Local Product 
space (BiLP).
\end{definition}
We also give a special name to
a BiLP with the isometries sketched above:
\begin{definition}[ALP]
A BiLP with these additional isometries 
(further characterized below) we will call an 
Auxilliary Product Space (ALP).
\end{definition}
The isometries are generated by $2n$ complex Abelian Killing
vectors that preserve all the BiLP geometric data. They
can be separated into two groups of $n$ Killing vectors
$\left\{k_{LA}\right\}$ and $\left\{k_{RA}\right\}$, $A=1\ldots n$,
satisfying the further requirement 
\beq
\imath_{k_{LA}}H = d\alpha_{LA}~,~~~
\imath_{k_{RA'}}H = -d\alpha_{RA'}~,
\eeq{kL}
where the one-form $\alpha$ is the dual of the killing 
vector $k$: $\alpha = g(k)$; this is equivalent to saying
that the Killing vectors are covariantly constant with respect
to the covariant derivative with torsion (\ref{3}).
Such Killing vectors are called Kac-Moody Killing
vectors. We require that each group 
of Killing vectors $k_L$ and $k_R$ 
by themselves span maximally isotropic subspaces, {\it i.e.,} that
\beq
g(k_{LA},k_{LB}) =0~,~~ 
g(k_{RA'},k_{RB'}) = 0~,
\eeq{null}
but that the inner product
\beq
h_{A'B} = g(k_{RA'},k_{LB})
\eeq{hAB}
is nondegenerate.

\begin{theorem}
Locally, any generalized K\"ahler manifold M is a quotient of an ALP
by its Kac-Moody isometries.
\end{theorem}

The quotient is performed by going to the orbits of the action of the Killing vector
and choosing a horizontal subspace by specifying a connection. In the particular
case with left and right Kac-Moody Killing vectors there is a corresponding left and
right connection given by
\beq
(\theta_R)^{A'}_\mu = g_{\mu\nu}  k^{\nu}_{LB}h^{BA'}~,~~~~~~~~~
(\theta_L)^A_\mu = h^{AB'}k^\nu_{RB'}  g_{\nu\mu}~,
\eeq{definsalxldoo}
where $h^{AB'}$ is the inverse of the nondegenerate matrix (\ref{hAB}).
The connections (\ref{definsalxldoo})
satisfy the following properties
\beq
\theta_L( k_{L})={\mathbf 1}~,~~~~
\theta_R( k_{R})={\mathbf 1}~,~~~~
\theta_L( k_{R})= 0~,~~~~
\theta_R( k_{L})= 0~.
\eeq{definsll2000}
Vectors in the horizontal subspaces are defined as the vectors lying in the
kernel of {\em both} $\theta_L$ and $\theta_R$. 

All geometric structures defined on the ALP can now be defined on the 
quotient space. As usual, forms are projected using the connection
so that the contraction with any vertical vector is zero,
whereas vectors are pushed forward to the quotient using
the bundle projection $\mathbf p$. When we project the complex structures
\beq
\tilde J_\pm(X) := {\mathbf p}_*\Big(J_\pm(X)-\theta_L(X)\cdot J_\pm(k_L)
-\theta_R(X)\cdot J_\pm(k_R)\Big)~\forall X~,
\eeq{Jquot}
it is clear that the projected complex structures $\tilde J_\pm$ do {\em not} commute
even though $J_\pm$ do.

Projecting the metric and $B$-field, we find
\ber
\tilde{g}  &=& g -h(\theta_L,\theta_R) - h(\theta_R,\theta_L)~,\cr\cr
\tilde{B} &=& B-h(\theta_L,\theta_R) + h(\theta_R,\theta_L)~.
\eer{qmet}  
It is straightforward to see that $\tilde{g}$ gives zero when contracted with any vertical
vector while this is not so for $\tilde{B}$. This (apparent) mystery is resolved by showing that
$\tilde{H} = d\tilde{B}$ can be written as
\beq
\tilde{H} = H -\theta_L \wedge \imath_{k_L} H
-\theta_R\wedge \imath_{k_R} H 
-\theta_L\wedge\theta_R\wedge \imath_{k_L}\imath_{k_R} H
\eeq{qH}
so that the geometrically meaningful object $\tilde{H}$ defined by $d\tilde{B}$ is
indeed well defined on the quotient. The quotients
$\tilde{g}$ and $\tilde{H}$ agree with
the expressions for the original generalized K\"ahler manifold.

\bigskip
\noindent{\bf\large Acknowledgement}:
\bigskip

\noindent We are grateful to the 2006 Simons Workshop
for providing the stimulating atmosphere where this work was initiated.
The work of UL was supported by EU grant (Superstring theory)
MRTN-2004-512194 and VR grant 621-2003-3454.
The work of MR was supported in part by NSF grant no.~PHY-0354776. 
The research of R.v.U. was supported by 
Czech ministry of education contract No. MSM0021622409.
The research of M.Z. was
supported by VR-grant 621-2004-3177

\end{document}